\documentclass[prl,twocolumn,showpacs,superscriptaddress,floatfix]{revtex4}
\usepackage{graphicx}
\usepackage{amssymb}
\usepackage{citesort}

\begin{document}

\title{The evolution of the Non-Fermi Liquid behavior of BaVS$_3$ under high pressure}

\author{N. Bari\v si\'c}
\affiliation{Institut de Physique de la mati\`{e}re complexe, EPFL, CH-1015 Lausanne, Switzerland}
\affiliation{Institut of Physics, Bijeni\v cka c. 46, Hr-10 000 Zagreb, Croatia}
\author{A. Akrap}
\author{H. Berger}
\author{L. Forr\'o} \affiliation{Institut de Physique de la mati\`{e}re complexe, EPFL, CH-1015 Lausanne, Switzerland}

\date{\today}

\begin{abstract}
Temperature, pressure, and magnetic field dependencies of the resistivity of BaVS$_3$ were measured above the critical pressure of $p_{cr}$=2 GPa, which is associated with the zero temperature insulator-to-metal (MI) transition. The resistivity exhibits the $T^n$ temperature dependence below $T_g\approx$15 K, with $n$ of 1.5 at $p_{cr}$, which increases continuously with pressure towards 2. This is interpreted as a crossover from non-Fermi (NFL) to Fermi-liquid (FL) behavior. Although the spin configuration of the $e_g$ electrons influences the charge propagation, the NFL behavior is attributed to the pseudogap that appears in the single particle spectrum of the $d_z^2$ electrons related to large quasi-one dimensional (Q-1d) 2$k_F$-CDW fluctuations. The non-monotonic magnetic field dependence of $\Delta$$\rho$/$\rho$ reveals a characteristic field $B_0\approx$12 T attributed to the full suppression of the pseudogap.

\end{abstract}

\pacs{71.10.Hf, 71.30.+H, 72.80.Ga}

\maketitle

BaVS$_3$ is a strongly correlated electron system which combines the properties of quasi-one dimensional (Q-1d) conductors, heavy fermion systems and conventional transition metal compounds. This versatility places it in the focus of experimental \cite{KezsmarkiOPTICS,Fagot} and theoretical \cite{Lechermann} investigations. The major challenges are the understanding of the metal-insulator transition, the magnetic structure and the high-pressure Non-Fermi liquid (NFL) state of the compound.

V and S atoms in BaVS$_3$ form well separated 1d VS$_3$ chains. At high temperatures, it is a metal with its V $d^1$-electron shared between the broad Q-1d $d_z^2$ band and the narrow, but more isotropic $e_g$ band \cite{Mitrovic}. In X-ray scattering diffuse lines show up below 150 K as the precursors to the charge density wave transition (MI), which happens at $T_{MI}\sim$69 K \cite{Fagot}. The 2$k_F$ value deduced from the position of the diffuse lines indicates that, already at temperatures well above $T_{MI}$, half of the V-electrons are in the $d_z^2$ band, which gives a $1/4$ filled band. This commensurability of 4 represents a rather weak potential \cite{Giamarchi} and is thus not sufficient alone to account for the $1/4$ filling of the $d_z^2$ band. This is, in part, attributed to the large on-site coupling \cite{Lechermann} between the $d_z^2$ and $e_g$ electrons, which also tends to align their spins according to Hund's rule. It should be noted that due to the commensurability effect of the order 4, the charge density wave (CDW) may also be accompanied by spin density wave (SDW) fluctuations\cite{Giamarchi,Solyom}.

The site degeneracy of the $e_g$ states is lifted by the Jahn-Teller distortion below $T_S\sim$230K, which leaves one $e_g$ state per site below the Fermi level. As a result BaVS$_3$ is close to a two-band system with $1/2+1/2$ sharing of the electrons. Coulomb interactions in the narrow $e_g$ band are presumably comparatively large in contrast to the $d_z^2$ band. They are however not large enough to localize completely the $e_g$ electrons in the metallic phase, where the latter are believed to ensure the relatively large interchain conductivity $\sigma_a/\sigma_c\sim$1/3 \cite{Mihaly}.

Concerning the magnetic properties, the MI transition produces a cusp in the large, anisotropic Curie-Weiss magnetic susceptibility in pure BaVS$_3$, \cite{Mihaly}. Unlike the resistivity, the uniform magnetic susceptibility is dominated by the narrow band $e_g$ electrons \cite{Mihaly}. Thus we attribute the cusp in the susceptibility at the MI transition to the above mentioned Hund coupling of paramagnetic ($e_g$) and SDW ($d_z^2$) subsystems. Magnetic fluctuations were observed below $T_{MI}$ and associated with pre-transitional effects of the low temperature magnetic transition \cite{HIGHfield}. Below $T_X\sim$30 K the system exhibits a long-range magnetic order \cite{TxRef}. The ordering along the V-chains is ferromagnetic with a long helical modulation \cite{TxRef}. The magnetic periodicity transversely to the chains is incommensurate. Although the overall magnetic response is often named "antiferromagnetic" we feel that the term "avoided ferromagnetism" (aFM) gives a better description of the magnetic structure. It is worthy of note that all other systems from the same family \cite{Gauzzi,FagotNEKOMENZ,SULFURdeficient} are incommensurate and exhibit the pure FM order at low temperatures. That suggests that the aFM order below $T_X$ originates from an interplay between the Hund-coupled FM ($e_g$) order and the SDW ($d_z^2$) order associated with the commensurability.

At a critical pressure $p_{cr}$ of 2 GPa the $T_{MI}$ is suppressed and a NFL behavior with a power-law dependence of the resistivity ($\rho=\rho_0 + AT^n$) appears \cite{Forro}. The microscopic origin of this NFL is not clear. To clarify the role of the $d_z^2$ and $e_g$ electrons in this state we apply pressure and magnetic field to tune the interaction within and in-between these bands. The resistivity in BaVS$_3$ was thus systematically followed as a function of pressure, temperature and magnetic field. Based on these measurements, in this letter we give an interpretation of the low energy properties of BaVS$_3$, especially above $p_{cr}$.

Special care was taken to use only high quality single crystals, since the physics of BaVS$_3$ is quite sensitive to disorder. Crystals were grown by previously established Tellurium flux method \cite{Kuriyaki} and carefully characterized as explained elsewhere \cite{Hysteresis,Thesis}. Samples with a typical size of 2x0.2x0.2 mm$^3$ were mounted into a nonmagnetic self-clamped pressure cell. The pressure was monitored in situ using a calibrated InSb pressure gauge. Kerosene was used as the pressure transmitting medium. The resistivity was measured along the chain direction and the applied magnetic field is perpendicular to the chains.

\begin{figure}[tb]
\includegraphics[totalheight=7cm]{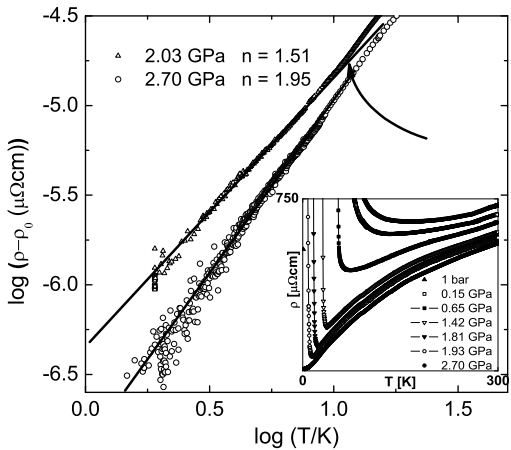}
\caption{ The low temperature (up to 40 K) part of the resistivity at 2.03 and 2.7 GPa, presented in the log-log plot, reveals the power-law dependence ($T^n$) of the resistivity below 15 K with coefficients $n=1.5$ and $n=1.95$, corresponding to the NFL and FL behaviors, respectively. The arrow denotes $T_g$ (see text). Inset: Temperature dependence of resistivity for various pressures showing the suppression of $T_{MI}$.}
\label{fig1}
\end{figure}

The $MI$ transition shifts to lower temperatures on increasing pressure (inset in Fig.\ref{fig1})\cite{KezsmarkiMAGRES,Hysteresis} reaching zero temperature at $p_{cr}$. Above this pressure, $p_{cr}$=1.97GPa, metallic behavior is observed in the whole temperature range studied (above 2 K). Up to 15-20 K the resistivity follows a $\rho=\rho_0 + AT^n$ behavior. The power law coefficient $n$ depends on the sample purity. We emphasize that in high quality samples $n=1.5$, canonical for a NFL, but it can be as low as 1.1 in impure samples \cite{Thesis,Forro}. Results in Fig.1 demonstrate that the NFL fit holds up to $T_g\approx$15 K. There are indications which suggest that $T_g$ is more than just a smooth cross-over temperature. $T_g$ is also the temperature at which the $T_{MI}$ collapses in the $p$-$T$ phase diagram (due to its crossing with $T_X$) and below which additional features appear in the close vicinity of $p_{cr}$ in pure samples \cite{Hysteresis}, e.g., magnetoresistance changes sign, hysteretic behavior in the conductivity, etc. We associate tentatively $T_g$ with the transition from the paramagnetic metallic to the ferromagnetic metallic state (PM-FM) \cite{Hysteresis,Thesis}.

The $n<2$ NFL power-law dependence of the resistivity of BaVS$_3$ below $T_g$ and close to $p_{cr}$ is similar to that found in some heavy fermion compounds, as pointed out previously \cite{Forro}. In those systems it is ascribed to the proximity of a quantum critical point (QCP) associated with the disappearance at $T=0$ of the magnetic order with the finite wave-vector $Q$. NFL behavior is predicted \cite{Millis} to occur in the resistivity of the heavy fermion systems when the energy of the magnetic fluctuations becomes less than $k_BT$. This corresponds to the adiabatic pseudogap regime \cite{Sunko}. In this regime, the conductivity is dominated by the finite $Q$ magnetic correlations in the band of the interacting conduction electrons \cite{Coleman}. In BaVS$_3$ the pseudogap which affects the Fermi surface of the $d_z^2$ band is associated, in the first place, with the intraband ($d_z^2$) 2$k_F$ charge fluctuations. As the CDW is coupled to the heavy lattice the adiabatic approximation is even more suitable than in the case of magnetic fluctuations. This is the basis of the analogy of the NFL behavior of our system to heavy fermions. The proposed mechanism should be strongly pressure dependent in BaVS$_3$, as Q-1d fluctuations are. It has to be emphasized that the relative weight of CDW and SDW fluctuations is not essential for the NFL behavior of the metallic conductivity, provided that $k_BT$ is sufficiently high to make the adiabatic approximation valid at least for CDW.

\begin{figure}[tb]
\includegraphics[totalheight=10cm]{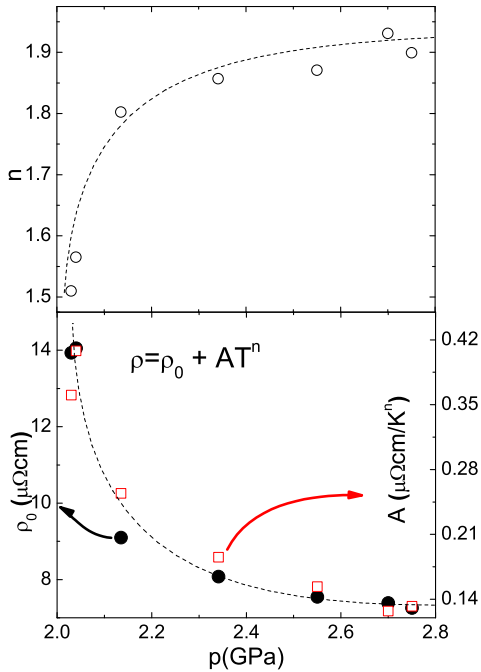}
\caption{ Pressure dependence of the (a) resistivity exponent $n$ (the dashed line is guide to the eye); (b) Pressure dependence of the prefactor $A$ (red squares) and the "residual" resistivity $\rho_0$ (black circles) as the results of the fits $\rho=\rho_0+AT^n$ in the temperature range $1.7<T<15$ K.}
\label{fig2}
\end{figure}

Fig.\ref{fig2} displays the $n$, $A$ and $\rho_0$ characteristic quantities deduced from resistivity measurements performed above $p_{cr}$. $n$ which is equal to 1.5$\pm$0.1 at $p_{cr}$=2.03 GPa, increases monotonically with pressure to 1.95$\pm$0.1 at 2.7 GPa, close to the Fermi-Liquid (FL) value of 2. The crossover in pressure from the NFL to the FL regime is clearly depicted through the coefficient $A$ and $\rho_0$, as well. $A(p)$ shown in Fig.\ref{fig2}b decreases when the material is driven away from $p_{cr}$. Similarly, strong pressure dependence of $n$ and $A$ is also found in the critical region e.g. in CeIn$_3$ \cite{Knebel}. The residual resistivity $\rho_0(p)$ which describes the behavior of the system in the low temperature limit also warrants attention. It is usually not possible to change by pressure or magnetic field the number and/or configuration of elastic scattering centers (impurities, defects) which give $\rho_0$ in the material. Nevertheless, the results presented in Fig.\ref{fig2}b clearly reveal a strong pressure dependence of $\rho_0$ and indicate that its origin is not in the scattering on impurities. The impurity contribution to $\rho_0=7.25 \mu\Omega cm$ is small and it is identified in the FL phase of BaVS$_3$ at 2.7 GPa. The scaling of $\rho_0$, $A$ and $n$ with pressure in Fig.\ref{fig2} shows that the same intrinsic physics is reflected in all these parameters.

These measurements corroborate with our view that close to the phase boundary, below $T_g$, substantial CDW fluctuations (i.e. short range lattice tetramerizations) still exist in BaVS$_3$ above $p_{cr}$. They give the NFL character of the resistivity. It should be noted that the scattering of the conduction electrons does not distinguish at sufficiently large $T$ between the static disorder or the low frequency dynamical (CDW/SDW fluctuations) \cite{Gogolin}. Since these fluctuations disappear with pressure, the NFL behavior switches to a standard FL one \cite {Gor´kov}.

\begin{figure}[tb]
\includegraphics[totalheight=10cm]{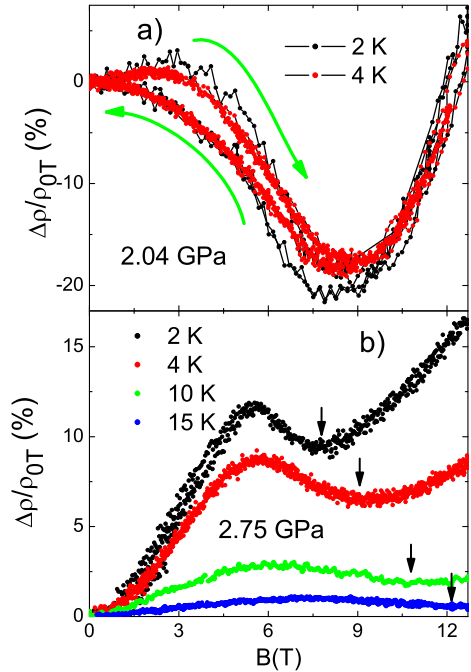}
\caption{ Isothermal magnetoresistance curve recorded at different temperatures and pressures. (a) The 2 and 4 K curves at 2.04 GPa. The direction of the evolution of the magnetic field is denoted by green arrows; (b) Curves taken at 2.75 GPa for several temperatures. The black arrows indicates the position of the dip at various temperatures.}
\label{fig3}
\end{figure}

The magnetic field $B$ strongly influences the resistivity of pure BaVS$_3$ below \cite{Hysteresis} and above $p_{cr}$. The magnetoresistance is radically different deep in the NFL regime and close to the FL state, as shown in Fig.\ref{fig3}. Just above $p_{cr}$, well within the NFL regime the magnetoresistance is negative at low fields (Fig.3a). On increasing field the domains of fluctuating CDW/SDW (tetramerized phase) are suppressed, either due to its direct coupling to the SDW component or due to the Zeeman-splitting of $k_F$ of the CDW \cite{Hysteresis}. Once these domains disappear the electrons which were involved in CDW/SDW are released, and improve the conductivity. This effect superimposes on the spin-dependent magnetoresistance (better alignment of $e_g$ spins with magnetic field) and results in the relatively strong negative magnetoresistance at 2 and 4 K. When all $d_z^2$ electrons involved in CDW formation are set free (at~9 T) the negative magnetoresistance saturates and reaches its highest absolute value. In this field range, due to the competition of the two periodicities associated with the commensurate CDW/SDW and the metallic FM, the magnetoresistance is associated with a hysteretic behavior\cite{BJELIS,Hysteresis}. After the full suppression of the commensurate CDW/SDW domains and the concomitant polarization of all spins (at~9 T) the hysteresis disappears. Above 9 T the orbital contribution to the magnetoresistance takes over and it tends towards positive values. On the other hand as the FL state is approached by increasing pressure the magnetoresistance changes its character. Already at 2.15 GPa (n=1.8, below $T_g$) the magnetoresistance is positive (not shown), and increases monotonically up to 9 T. However it does so with a slightly weaker slope compared to that at 2.7 GPa in the nearly FL state shown in Fig.3b, which we now elaborate upon. As in conventional metals, $\Delta$$\rho$/$\rho$ coming from orbital motion, satisfies Kohler's rule, that is $\Delta$$\rho$/$\rho$=f(B/$\rho$)(Fig.4). The fact that magnetoresistance taken at different temperatures scales together at low fields means that the underlying band structure, the carrier density, and the (constant) scattering rate do not change appreciably with the field. However, the difference in the Kohler plot with respect to conventional metals is that the magnetoresistivity starts to deviate from pure quadratic field dependence as the field is increased resulting in a dip at $\approx$9 T (for the curve taken at 4K). Such a behavior can be described along the lines discussed for magnetoresistance at lower pressures (neglecting the hysteresis) and for the NFL to FL crossover of the transport coefficients $n$, $A$ and $\rho_0$ (Fig.2). The positive contribution to magnetoresistance is predominately originating, like in conventional metals, from the circulation around the Fermi surface of the electrons available at $B=0$ T for a given pressure. The negative contribution, causing the deviation from simple quadratic behavior is associated here, in particular, to the field suppression of the pseudogap from the Fermi surface. This liberates the remaining electrons and consequently decreases the resistivity. Beyond the dip, when the carrier density saturates at the maximal value, only the positive Kohler-like term is present. This can include the slow variation of the scattering rate of the $d_z^2$ electrons on the spin disorder of the $e_g$ electrons with the field. The main lines of the magnetoresistance behavior are thus brought into qualitative agreement with our interpretation of the NFL to FL crossover and with the suggested pressure dependences of $T_{MI}$ and $T_X$ (replaced by $T_g$ at pressure above $p_{cr}$).

Inspection of Figs.\ref{fig3}a and \ref{fig3}b reveals that $B_0$ is weakly dependent on pressure and increases somewhat with temperature. Between 2K and 15K $B_0$ changes from 8 to 12T respectively (see Fig.\ref{fig3}b). This corroborates the assumption that FM is present below 15K and indicates that the $T\approx$0 local Weiss field is around 3T.

The magnetoresistance data of Fig.\ref{fig3} show that there is a strong interaction between the $d_z^2$ and $e_g$ electrons. Although the spin configuration of the $e_g$ electrons influences substantially the propagation of the charges, it is not the interaction which gives the NFL behavior at pressures close to the critical one. This behavior is solely attributed to the scattering on CDW fluctuations.

\begin{figure}[tb]
\includegraphics[totalheight=7cm]{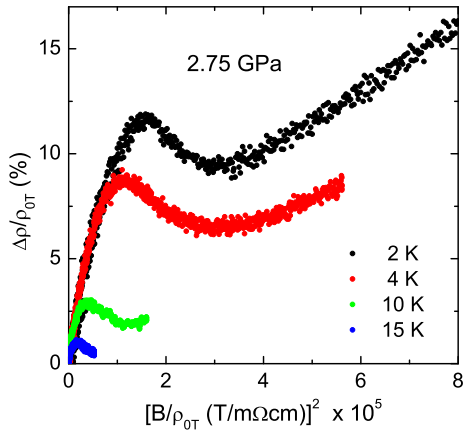}
\caption{ Kohler's plot of the isothermal magnetoresistance curves recorded at different temperatures and pressures, presented in Fig.\ref{fig3}. At magnetic fields below 9T, the magnetoresistance satisfies Kohler's law.}
\label{fig4}
\end{figure}

In conclusion we claim that BaVS$_3$ represents a special class of multiband NFL materials. In some heavy fermion systems the interactions are strong within the conduction band, hybridized in turn with the narrow f-state carrying the spin. These strong interactions induce the correlations which are reflected in the overall behavior of the system \cite{Coleman,Doniach}. It is however difficult to tune the strong interactions in the conduction band over a broad range by pressure or magnetic field. In BaVS$_3$ the weak interactions within the 2 eV wide 1d $d_z^2$ band generate the corresponding 1d CDW/SDW fluctuations. Just above $p_{cr}$, the long-range ordered CDW/SDW state is precluded by the pressure but the Fermi surface is still pseudo-gapped with hot spots analogous to the heavy fermion systems \cite{Coleman}. In both cases the result is the NFL behavior. Eventually high pressure removes the 1d fluctuations entirely and establishes the FL behavior. The present magnetoresistance measurements mimic the NFL to FL cross-over. Further investigations of the picture proposed here are nevertheless required to probe the CDW fluctuations and magnetic correlations at high pressures more directly than the resistivity and magnetoresistivity measured here.

Useful discussions with S. Bari\v{s}i\'{c}, J.P. Pouget, I. Kup\v{c}i\'{c}, T. Feher and A. Smontara are acknowledged. This work was supported by the Croatian and the Swiss National Foundation for Scientific Research and its pools MoSES 035-0352826-2848 and "MaNEP" respectively.

\end{document}